\input phyzzx\def\e{\adveq\eqno{\rm (\chapterlabel\the\equanumber)}}

\def\adveq{\global\advance\equanumber by 1}
\def\myeq{{\rm \chapterlabel\the\equanumber}}
\def\rarrow{\rightarrow}

\def\twoline#1#2{\displaylines{\qquad#1\hfill(\adveq\myeq)\cr\hfill#2
\qquad\cr}}

\def\semidirect{\mathrel{\raise0.04cm\hbox{${\scriptscriptstyle |\!}$
\hskip-0.175cm}\times}}

\def\mod{\mathop{\rm mod}\nolimits}

\def\ref#1{$^{[#1]}$}

\def\pr#1{#1^\prime}
 
\def\r#1{$[\rm#1]$} 
\def\twidle{\tilde}

\def\e{\adveq\eqno{\rm (\chapterlabel\the\equanumber)}}

\def\adveq{\global\advance\equanumber by 1}
\def\myeq{{\rm \chapterlabel\the\equanumber}}
\def\rarrow{\rightarrow}

\def\twoline#1#2{\displaylines{\qquad#1\hfill(\adveq\myeq)\cr\hfill#2
\qquad\cr}}

\def\semidirect{\mathrel{\raise0.04cm\hbox{${\scriptscriptstyle |\!}$
\hskip-0.175cm}\times}}

\def\mod{\mathop{\rm mod}\nolimits}

\def\ref#1{$^{[#1]}$}

\def\pr#1{#1^\prime}
 
\def\r#1{$[\rm#1]$} 
\def\twidle{\tilde}

\date{March, 2010}
\date{March, 2010}
\titlepage
\title{Nonabelian Parafermions and their Dimensions}
\author{Roman Dovgard and Doron Gepner}
\line{\hfill Department of Particle Physics\hfill}
\line{\hfill Weizmann Institute\hfill}
\line{\hfill Rehovot 76100, Israel\hfill}
\abstract
We propose a generalization of the Zamolodchikov-Fateev parafermions which 
are abelian, to nonabelian groups. The fusion rules are given by the tensor product
of representations of the group. Using Vafa equations we get the allowed
dimensions of the parafermions. We find for simple groups that the dimensions are
integers. For cover groups of simple groups, we find, for $n.G.m$, that the dimensions are the
same as $Z_n$ parafermions. Examples of integral parafermionic systems are 
studied in detail.  
\endpage

Conformal field theory in two dimensions has been a source of numerous results owing to
its solvability and its rich structure. It has been successfully applied to statistical mechanics and string theory. 

One line of such ideas is a generalization of the Ising model to so called parafermions, first put
forwards by Zamolodchikov and Fateev
\REF\ZF{A.B. Zamolodchikov and V.A. Fateev, Sov. Phys. JETP 62 (1985) 215.}\r\ZF.
These are analytic currents with nonintegral spin.
The known examples, so far, are based on the cyclic group $Z_n$, or products of cyclic groups,
that is, abelian groups. The parafermions appear in nature as fixed points of some models of magnets,
e.g., the Andrews Baxter Forrester models 
\REF\ABF{G.E. Andrews, R.J. Baxter and P.J. Forrester, J. Stat. Phys. 35 (1984) 193;
D.A. Huse, Phys. Rev. B 30 (1984) 3908.}\r\ABF\ 
and $Z_n$ clock models \r\ZF. Also, they are vital components in
string compactification, since they are closely related to $N=2$ superconformal theories, 
yielding solvable realistic string theories, featuring for example, in the 
models of ref. 
\REF\Gep{D. Gepner, Nucl. Phys. B 296 (1988) 757.}\r\Gep.

Our idea is to generalize the parafermions to nonabelian groups. We can write a parafermionic system 
for any group $G$. This we do by assuming that the currents fall into representations of the group
$G$. I.e., we have a parafermionic multiplet for each representation of the group, $G$, and
for each vector of the representation. We then postulate that the OPE of the parafermionic system obey
the group symmetry. This is a straight forwards generalization of the notion of parafermions and, in the
abelian case, it gives the usual results of Zamolodchikov and Fateev.

To be specific, assume that $I$ ranges over the representations of the group $G$ and that the index $i$ ranges
over the vectors in each representation. We introduce a parafermion which is a field $\psi^I_i(z)$, which is
an a holomorphic field of dimension $\Delta_I$. Let $f^{IJK}_{ijk}$ denote the Clebsh-Gordon coefficient
of the group. I.e., for each element of the group $g\in G$ we have the relation,
\def\pr#1{{#1^\prime}}
$$\sum_{\pr i,\pr j,\pr k} \Phi^I_{i\pr i}(g) \Phi^J_{j\pr j}(g) \Phi^K_{k\pr k}(g) f^{IJK}_{\pr i\pr j\pr k}=  
f^{IJK}_{ijk},\e$$
where $\Phi^I_{i\pr i}(g)$ is a matrix in the $I$th representation of the group $G$.

We further define, 
$$f^{IJ1}_{ij1}=b^{IJ}_{ij},\e$$
and
$$\sum_k b^{K\bar K}_{\pr k k} f^{k(IJ\bar K)}_{ij}=f_{ij\pr k}^{IJK}.\e$$

We then postulate the following operator product expansions (OPE) for the parafermions,
$$\twoline{\psi^I_i(z) \psi^J_j(w)={b_{ij}^{IJ} C_I\over (z-w)^{2\Delta_I}}+}{
\sum_{k,K} f_{ij}^{k(IJK)} (z-w)^{-\Delta_I-\Delta_J+\Delta_K}
\left[ {\twidle C}^{IJ}_K \psi^K_k(w)+{\twidle{\twidle C}}^{IJ}_K (z-w) \partial \psi^K_k(w)
\right]+{\rm h.o.t.},}$$
where ${\rm h.o.t.}$ stands for higher order terms. The constants $C$, $\twidle C$ and $\twidle{\twidle C}$ are
determined by the associativity of the OPE above, once the dimensions $\Delta_I$ are determined.
We postulate also $\psi^1_1(z)=T(z)$, the stress tensor, $\Delta_1=2$, $C_1=c/2$, where $c$ is the central charge.
Thus the algebra contains the Virasoro algebra and we demand, accordingly, that, 
$\psi^I_i(z)$ is a primary field,
${\twidle C}^{1 I}_I=\Delta_I$,
and, ${\twidle{\twidle C}}^{1 I}_I=1$.

The fusion rules, i.e., the way operators fuse in the operator product expansion, are then given simply
by the tensor product algebra of the representations of the group. This is easily calculated in specific
examples by means of the character tables of specific groups. 
Denote by $\chi_I(g)$ the character in the representation $I$ of the group, $g\in G$,
$$\chi_I(g)=\sum_i \Phi^I_{i i} (g).\e$$
Then the fusion coefficients of the parafermions are given by,
$$N^K_{IJ}={1\over O(G)} \sum_{g\in G} \chi_I(g) \chi_J(g) \chi_K(g)^*,\e$$
which is reminiscent of the Verlinde formula
\REF\Verlinde{E. Verlinde, Nucl. Phys. B 300 (1988) 360.}\r\Verlinde.
Here, $O(G)$ is the order of the group. 

Thus, we can use Vafa's equations 
\REF\Vafa{C. Vafa, Phys. Lett. B 206 (1988) 360.}\r\Vafa\
to calculate
the dimensions of the parafermions, which are found up to some arbitrary integer multiplicative factor.
Denote by 
$$\alpha_I=e^{2\pi i\Delta_I}.\e$$
Then Vafa equations are
$$(\alpha_I \alpha_J\alpha_K\alpha_L)^{N_{IJKL}}=
\prod_R \alpha_R^{N_{IJKL;R}},\e$$
where 
$$N_{IJKL}=\sum_R N_{IJR} N^R_{KL},\e$$
and
$$N_{IJKL;R}=N_{IJR} N^R_{KL}+N_{IKR} N^R_{JL}+N_{ILR} N^R_{JK},\e$$
where we define $N_{IJK}=N^{\bar K}_{IJ}$.

When $G$ is abelian, we are back in the case of Zamolodchikov and Fateev. For a $G=Z_N$ group, 
we denote the $I$ parafermion for the 
representation $\Phi_I(e^{2  \pi i r/N})=e^{2\pi i r I/N}$, for any $I$ and $r$ modulo $N$.
The dimension of the $I$th parafermion is $\Delta_I$ and $\Delta_I=\Delta_{N-I}$ since it is the
complex conjugate field. We find from eq. (6) that the structure constant is $N_{IJ}^K=1$ if $K-I-J=0\mod N$ and
is zero otherwise.
Here Vafa's equations become,
$$\Delta_I+\Delta_J+\Delta_K+\Delta_L=\Delta_{K+L}+\Delta_{K+I}+\Delta_{K+J} \mod Z,\e$$
where $I+J+K+L=0\mod N$ are any. This equation, already appears in ref. \r\ZF, eq. (A4) there, derived from the mutual 
semilocality.
This eq. (11) implies, in particular, by taking $I=J=1$, that 
$$2\Delta_{K+1}-\Delta_K-\Delta_{K+2}=\beta \mod Z,\e$$
where $\beta=2 \Delta_1-\Delta_2$. Thus, $\Delta_K=-\beta K^2/2\mod Z$ is the unique solution to eq. (11), which 
satisfies, $\Delta_0=$integer and $\Delta_1=\Delta_{N-1}$.
It follows that
$$\Delta_I=M_I+m I^2/(s N),\e$$
where $M_I$ and $m$ are arbitrary integers and $s=1$ for odd $N$ and $s=2$ for even $N$. We set $\Delta_r=\Delta_{N-r}$.
Thus, this method is consistent with the known abelian case.

Thus, for each group we simply substitute the characters into Vafa equations to find the dimensions of the
parafermions.

Let us introduce some basic notions of group theory. The group $G$ is called simple if the only normal subgroups
are itself or the trivial one. An automorphism is a one to one and onto map $\sigma:G\rarrow G$ such that
$$\sigma(gh)=\sigma(g)\sigma(h),\e$$
where $g,h\in G$. An internal automorphism is the map,
$$\sigma_h(g)=h g h^{-1},\e$$
where $h$ is a fixed element of the group. We denote the automorphism group by ${\rm Aut}(G)$, which is a group under
decompositions. We denote by ${\rm Int}(G)$ the internal automorphism subgroup of ${\rm Aut}(G)$, which is a 
normal subgroup. The outer automorphism group, ${\rm Out}(G)$ is defined as the quotient group,
$${\rm Out}(G)={{\rm Aut}(G) \over {\rm Int }(G)}.\e$$
The group $G$ itself is a subgroup of ${\rm Aut}(G)$ by identifying it with ${\rm Int}(G)$
(we assume that $G$ is centerless, see below. If we denote the center by $Z(G)$ then,
${\rm Int}(G)\approx G/Z(G)$).
A group, $H$, is called almost simple, if there exists a simple group $G$ such that
$$G\subset H \subset {\rm Aut}(G).\e$$
We call $H$ a cover group by automorphism of the group $G$.

We call the parafermion system integral if the only solution to Vafa equation is that all the dimensions are
integral. Our result about this can be phrased as:

{\rm Conjecture (1):} The parafermion system of the group $H$ is integral if and only if $H$ is a nonabelian
almost simple group or the trivial group.

Another type of cover group is an extension by a center. The center of a group $H$ is the subgroup of 
elements that commute with every group member. The center of the group $H$ is a normal subgroup denoted by $Z(H)$, 
$h\in Z(H)$ if and only if, $gh=hg$ for all $g\in H$. Of course the center of a nonabelian
simple group is trivial. If $H$ is a group such that $G=H/Z(H)$, where $G$ is a simple group, we call
$H$ a cover group of $G$ by a center. Other type of center group is mixed both by an automorphism and by
a center, where $W=H/Z(H)$ is an almost simple group of the simple group $G$, $G\subset W \subset {\rm Aut}(G)$.
We denote such a cover group as $H=n.G.m$, where the center is a $Z_n$ group and the outer automorphism
is a $Z_m$ group, i.e., the quotient of the almost simple group to its simple subgroup is a $Z_m$ group. 
Of course, the outer automorphisms group can be nonabelian (for a simple group it is always
solvable). 

Now, suppose that $H$ is a $Z_n$ cover group of an almost simple group $G$, $G=H/Z(H)$, where $Z(H)\approx Z_n$. 
Let $I$ be 
an irreducible representation of $H$. By Schure's lemma, since the center commutes with all elements of $H$, it
is an irreducible representation of $Z_n$. These representations are denoted by their charge modulo $n$.
So let $d_I\mod n$ be the charge of the $I$th representation.
Our result can then be quoted as:
 
{\rm Conjecture (2):} The parafermion system of a cover group of the type $H=n.G.m$, where $G$ is a nonabelian
simple group, gives the same dimensions as $Z_n$ parafermions. I.e., the dimension of the $I$th  
representation is, 
$$\Delta_I=M_I+m d_I^2/(s n),\e$$
where $s=1$ ($s=2$) for odd (even) $n$, and $M_I$ and $m$ are some integers.

The same applies to a center which is a product of cyclic groups. One just adds up the dimensions.

We can give a sort of proof for this conjecture. Let $d_I=0$ then it is an irreducible representation of $H$, where 
$G=H/Z(H)$, $G$ is almost simple. Since the representation $I$ acts trivially on the center, 
it can be lifted to a representation of $G$. 
Thus, by conjecture (1), the corresponding parafermion, $\psi^I_i$,
has integral dimension, and we can include it in the chiral algebra. So, all representations with
$d_I=0$ can be collected into an extended chiral algebra. Similarly, all representations with the same $d_I\neq0$
can be collected into one representation of this extended algebra. Thus, the theory is, in fact, a $Z_n$ parafermionic
system with this extended algebra and, using Vafa's equations, conjecture (2) follows.  

The characters tables for the simple groups and their cover groups are enumerated in the atlas of finite
groups \REF\Atlas{Atlas of Finite Groups, J.H. Conway et al, Oxford Univ. press (2003).}
\r\Atlas. We wrote a computer program to solve Vafa's equations for each case.
This way we found the dimensions for all the simple groups with up to $14$ conjugacy classes, along with most
of their cover groups. We find that, indeed, conjectures (1) and (2) are obeyed and for any group of the type
$n.G.m$, where $G$ is simple, the dimensions are as for $Z_n$ parafermions in accordance with conjecture (2). 

Our objects here is to study the nonabelian parafermions. There is one special case which is simpler to analyze.
This is the case where all the dimensions are integral. We have such a solution for any group. Of course, if
the group is almost simple and nonabelian, this is the only solution. For simplicity we assume that all the
dimensions are two, $\Delta_I=2$, for all the representations $I$. Then, the OPE, eq. (4), assumes the form,
$$\psi_i(z) \psi_j(w)={b_{ij} c/2\over (z-w)^4}+f_{ij}^k \left[ {2\psi_k(w) \over (z-w)^2}+{\partial\psi_k(w)\over
(z-w)}\right],\e$$
up to regular terms, where for simplicity of notation we denote $\psi^I_i(w)$ by one index
$\psi_i(w)$. Of course we can have such an algebra for any $b_{ij}$ and any $f_{ij}^k$ provided they obey
the associativity of the OPE. To check the latter, it is convenient to write the algebra in modes,
$$\psi_i(w)=\sum_n L_i^n w^{-n-2},\e$$
and we get the commutator relations,
$$[L_i^n,L_j^m]=b_{ij} (c/12) (n^3-n)\delta_{n+m,0}+\sum_k f_{ij}^k (n-m) L_k^{n+m}.\e$$
We take, of course, $\psi_0(w)=T(w)$, the stress tensor, and $b_{00}=f_{00}^0=1$.
The associativity of the OPE is equivalent to the Jacobi identity,
$$[A,[B,C]]+[B,[C,A]]+[C,[A,B]]=0,\e$$
which, in this case, is obeyed if and only if the following relations hold,
$$\sum_k b_{ik} f^k_{jl}=f_{ijl},\e$$
and $f_{ijl}$ is fully symmetric in all indices, along with
$$\sum_l f_{ik}^l f_{jl}^f=\sum_l f_{ij}^l f_{lk}^f,\e$$
for all $i,j,k,f$. With these relations the OPE is associative and we have a conformal field theory with this
extended algebra. In addition, unitarity requires that
$$(f_{ij}^k)^*=f_{\bar i \bar j}^{\bar k}.\e$$

Let us consider the case where $G=Z_N$, that is, abelian parafermions, with the algebra eq. (21). Here we
assume that the representation weights are $i$, $j$ and $k$, defined modulo $N$, and that,
$$f_{i,j}^{i+j}\neq 0,\qquad f^k_{i,j}=0 {\ \rm where\ } k\neq i+j\mod N\e$$
Also,
$$b_{i,j}=\delta_N(i+j),\e$$
where $\delta_N(x)=1$ if $x=0\mod N$ and is zero otherwise. The unitarity requirement 
becomes,
$$f_{N-i,N-j}^{2 N-i-j}=(f_{i,j}^k)^*.\e$$
The solution to the structure constants obeying eqs. (23-28) can be seen to be given by
$$f_{i0}^i=1,\e$$
$$f_{i,j}^{i+j}={a_i a_{i+1}\ldots a_{i+j-1} \over a_1 a_2\ldots a_{j-1} },\e$$
where 
$$a_i=f_{i,1}^{i+1},\e$$
and
$$a_N=a_{N-1}=1,\qquad a_r=e^{i\phi_r},\e$$
where $\phi_r$ is a real phase obeying,
$$\phi_r=\phi_{r+N}=\phi_{N-1-r}.\e$$
This can be seen to be the most general solution, obeying the Jacobi identity and unitarity for any 
values of the real parameters $\phi_r$, $r=1,2,\ldots,[(N-2)/2]$, where $[x]$ is the 
smallest integer bigger or equal to $x$.
Thus we have a multiparameter algebra depending on $[(N-2)/2]$ real angles. 

To study the unitarity of these field theories we can use the Kac determinant method. This is left for
future work. There is one simple case, though. This is when all the angles vanish,
$$f_{ij}^{i+j}=1,\e$$
for all $i$ and $j$. The algebra then assumes the form
$$[L_n^i,L_m^j]={c\over 12} (n^3-n)\delta_{n+m}\delta_N(i+j)+(n-m)L_{n+m}^{i+j},\e$$
where $n,m$ are integers, and $i,j$ integers modulo $N$.
We can define a new basis for the algebra as,
$$M_n^r={1\over N} \sum_{s=0}^{N-1} e^{2\pi i s r/N} L_n^s,\e$$
where $r$ is defined modulo $N$.
It is easy then to see that the $M$'s form a set of commuting Virasoro algebras, 
all with the same central charge,
$$[M_n^i,M_s^j]=\delta_{i,j}\left[ {c\over 12 N} (n^3-n)\delta_{n+s}+(n-s) M_{n+s}^i\right]
.\e$$
So the theory is a tensor product of Virasoro algebras with the central charge $c/N$. 
We conclude that the unitary minimal models of this algebra are for $c<N$ and the central 
charge and dimensions are
given by,
$$c=N\left(1-{6\over m(m+1)}\right),\qquad m=2,3,4,\ldots,\e$$
$$h=\sum_{i=1}^N {(m p_i-(m+1) q_i)^2-1 \over 4 m (m+1)},\e$$
where $1\leq p_i\leq m$ and $1\leq q_i\leq m-1$ are arbitrary integers.

When the angles in the algebra do not vanish, the theories are nontrivial and a preliminary study of the Kac
determinant shows that they are no longer a tensor product of simpler objects.

The possible applications of this work are, as indicated in the introduction, its immediate importance
to both statistical systems and string theory. For the latter, many parafermionic systems can be completed
to an $N=2$ superconformal field theories, which can then yield space--time supersymmetric string theories 
in four dimensions \r\Gep. It is an obvious step to study the spectrum and realism of such string theories.
In particular, if $\Delta_I<3/2$ we can define the superconformal stress tensor by,
$$G_+(z)=\sqrt{2 c+2\over 3}  \psi_i^I(z) :e^{i\gamma\phi(z)}:,\e$$
$$G_-(z)=\sqrt{2 c+2\over 3 } \psi_{\bar i}^{\bar I}(z) :e^{-i\gamma\phi(z)}:,\e$$
where $\phi(z)$ is a canonical free boson and $\gamma^2/2+\Delta_I=3/2$, we normalized $b^{I\bar I}_{i\bar i} C_I=1$,
and $c$ is the central charge of the parafermionic theory.
The $N=2$ superconformal algebra will then be obeyed provided the central charge and the dimension are related by
$$c={2\Delta_I\over 3-2\Delta_I},\e$$
which is obeyed in many cases.

\refout
\bye